\documentstyle[twoside]{article}

\catcode`\@=11
\long\def\@makefntext#1{
\protect\noindent \hbox to 3.2pt {\hskip-.9pt
$^{{\eightrm\@thefnmark}}$\hfil}#1\hfill}       %CAN BE USED

\def\@makefnmark{\hbox to 0pt{$^{\@thefnmark}$\hss}}    %ORIGINAL

\def\ps@myheadings{\let\@mkboth\@gobbletwo
\def\@oddhead{\hbox{}
\rightmark\hfil\eightrm\thepage}
\def\@oddfoot{}\def\@evenhead{\eightrm\thepage\hfil
\leftmark\hbox{}}\def\@evenfoot{}
\def\sectionmark##1{}\def\subsectionmark##1{}}

\oddsidemargin=\evensidemargin
\newcounter{sectionc}\newcounter{subsectionc}\newcounter{subsubsectionc}
\renewcommand{\section}[1] {\vspace{12pt}\addtocounter{sectionc}{1}
\setcounter{subsectionc}{0}\setcounter{subsubsectionc}{0}\noindent
    {\tenbf\thesectionc. #1}\par\vspace{5pt}}
\renewcommand{\subsection}[1] {\vspace{12pt}\addtocounter{subsectionc}{1}
    \setcounter{subsubsectionc}{0}\noindent
    {\bf\thesectionc.\thesubsectionc. {\kern1pt \bfit #1}}\par\vspace{5pt}}
\renewcommand{\subsubsection}[1] {\vspace{12pt}\addtocounter{subsubsectionc}{1}
    \noindent{\tenrm\thesectionc.\thesubsectionc.\thesubsubsectionc.
    {\kern1pt \tenit #1}}\par\vspace{5pt}}

\newcounter{appendixc}
\newcounter{subappendixc}[appendixc]
\newcounter{subsubappendixc}[subappendixc]
\renewcommand{\thesubappendixc}{\Alph{appendixc}.\arabic{subappendixc}}
\renewcommand{\thesubsubappendixc}
    {\Alph{appendixc}.\arabic{subappendixc}.\arabic{subsubappendixc}}

\renewcommand{\appendix}[1] {\vspace{12pt}
        \refstepcounter{appendixc}
        \setcounter{figure}{0}
        \setcounter{table}{0}
        \setcounter{lemma}{0}
        \setcounter{theorem}{0}
        \setcounter{corollary}{0}
        \setcounter{definition}{0}
        \setcounter{equation}{0}
        \renewcommand{\thefigure}{\Alph{appendixc}.\arabic{figure}}
        \renewcommand{\thetable}{\Alph{appendixc}.\arabic{table}}
        \renewcommand{\theappendixc}{\Alph{appendixc}}
        \renewcommand{\thelemma}{\Alph{appendixc}.\arabic{lemma}}
        \renewcommand{\thetheorem}{\Alph{appendixc}.\arabic{theorem}}
        \renewcommand{\thedefinition}{\Alph{appendixc}.\arabic{definition}}
        \renewcommand{\thecorollary}{\Alph{appendixc}.\arabic{corollary}}
        \renewcommand{\theequation}{\Alph{appendixc}.\arabic{equation}}
        \noindent{\tenbf Appendix \theappendixc #1}\par\vspace{5pt}}
\newcommand{\subappendix}[1] {\vspace{12pt}
        \refstepcounter{subappendixc}
        \noindent{\bf Appendix \thesubappendixc. {\kern1pt \bfit #1}}
    \par\vspace{5pt}}
\newcommand{\subsubappendix}[1] {\vspace{12pt}
        \refstepcounter{subsubappendixc}
        \noindent{\rm Appendix \thesubsubappendixc. {\kern1pt \tenit #1}}
    \par\vspace{5pt}}

\newcommand{\textlineskip}{\baselineskip=13pt}
\newcommand{\smalllineskip}{\baselineskip=10pt}

\def\eightcirc{
\begin{picture}(0,0)
\put(4.4,1.8){\circle{6.5}}
\end{picture}}
\def\eightcopyright{\eightcirc\kern2.7pt\hbox{\eightrm c}}

\newcommand{\copyrightheading}[1]
    {\vspace*{-2.5cm}\smalllineskip{\flushleft
    {\footnotesize International Journal of Modern Physics A, #1}\\
    {\footnotesize $\eightcopyright$\, World Scientific Publishing
     Company}\\
     }}

\def\abstracts#1#2#3{{
    \centering{\begin{minipage}{4.5in}\baselineskip=10pt\footnotesize
    \parindent=0pt #1\par
    \parindent=15pt #2\par
    \parindent=15pt #3
    \end{minipage}}\par}}

\newcommand{\bibit}{\nineit}

\renewenvironment{thebibliography}[1]
    {\frenchspacing
     \ninerm\baselineskip=11pt
     \begin{list}{\arabic{enumi}.}
    {\usecounter{enumi}\setlength{\parsep}{0pt}
     \setlength{\leftmargin 12.7pt}{\rightmargin 0pt} %FOR 1--9 ITEMS
     \setlength{\itemsep}{0pt} \settowidth
    {\labelwidth}{#1.}\sloppy}}{\end{list}}

\newcounter{itemlistc}
\newcounter{romanlistc}
\newcounter{alphlistc}
\newcounter{arabiclistc}

\newcommand{\fcaption}[1]{
        \refstepcounter{figure}
        \setbox\@tempboxa = \hbox{\footnotesize Fig.~\thefigure. #1}
        \ifdim \wd\@tempboxa > 5in
           {\begin{center}
        \parbox{5in}{\footnotesize\smalllineskip Fig.~\thefigure. #1}
            \end{center}}
        \else
             {\begin{center}
             {\footnotesize Fig.~\thefigure. #1}
              \end{center}}
        \fi}

\newcommand{\tcaption}[1]{
        \refstepcounter{table}
        \setbox\@tempboxa = \hbox{\footnotesize Table~\thetable. #1}
        \ifdim \wd\@tempboxa > 5in
           {\begin{center}
        \parbox{5in}{\footnotesize\smalllineskip Table~\thetable. #1}
            \end{center}}
        \else
             {\begin{center}
             {\footnotesize Table~\thetable. #1}
              \end{center}}
        \fi}

\def\@citex[#1]#2{\if@filesw\immediate\write\@auxout
    {\string\citation{#2}}\fi
\def\@citea{}\@cite{\@for\@citeb:=#2\do
    {\@citea\def\@citea{,}\@ifundefined
    {b@\@citeb}{{\bf ?}\@warning
    {Citation `\@citeb' on page \thepage \space undefined}}
    {\csname b@\@citeb\endcsname}}}{#1}}

\newif\if@cghi
\def\cite{\@cghitrue\@ifnextchar [{\@tempswatrue
    \@citex}{\@tempswafalse\@citex[]}}
\def\citelow{\@cghifalse\@ifnextchar [{\@tempswatrue
    \@citex}{\@tempswafalse\@citex[]}}
\def\@cite#1#2{{$\null^{#1}$\if@tempswa\typeout
    {IJCGA warning: optional citation argument
    ignored: `#2'} \fi}}

\def\pmb#1{\setbox0=\hbox{#1}
    \kern-.025em\copy0\kern-\wd0
    \kern.05em\copy0\kern-\wd0
    \kern-.025em\raise.0433em\box0}

\def\fnt#1#2{\footnotetext{\kern-.3em
    {$^{\mbox{\scriptsize #1}}$}{#2}}}

\def\fpage#1{\begingroup
\voffset=.3in
\thispagestyle{empty}\begin{table}[b]\centerline{\footnotesize #1}
    \end{table}\endgroup}

\def\runninghead#1#2{\pagestyle{myheadings}
\markboth{{\protect\footnotesize\it{\quad #1}}\hfill}
{\hfill{\protect\footnotesize\it{#2\quad}}}}
\font\tenrm=cmr10
\font\tenit=cmti10
\font\tenbf=cmbx10
\font\bfit=cmbxti10 at 10pt
\font\ninerm=cmr9
\font\nineit=cmti9

\font\eightrm=cmr8

\def\qed{\hbox{${\vcenter{\vbox{            %HOLLOW SQUARE
   \hrule height 0.4pt\hbox{\vrule width 0.4pt height 6pt
   \kern5pt\vrule width 0.4pt}\hrule height 0.4pt}}}$}}

\def\bbox#1{%
\leavevmode\text{%
\textfont0 \the\textfont\bffam
\scriptfont0 \the\scriptfont\bffam
\scriptscriptfont0 \the\scriptscriptfont\bffam
\@temptokena\everymath \boldmath \everymath\@temptokena
{$\m@th\relax#1$}%
}%
}
\def\bbox#1{%
\relax\ifmmode
\mathchoice
{{\hbox{\boldmath$\displaystyle#1$}}}%
{{\hbox{\boldmath$\textstyle#1$}}}%
{{\hbox{\boldmath$\scriptstyle#1$}}}%
{{\hbox{\boldmath$\scriptscriptstyle#1$}}}%
\glb@settings
\else
\mbox{#1}%
\fi
}

\begin{document}

\runninghead{Real-time Nonequilibrium Dynamics in Hot QED Plasmas}
{Real-time Nonequilibrium Dynamics in Hot QED Plasmas}

\normalsize\textlineskip

\thispagestyle{empty}

\setcounter{page}{1}

\copyrightheading{}         %{Vol. 0, No. 0 (1993) 000--000}

\vspace*{0.88truein}

\fpage{1} \centerline{\bf REAL-TIME NONEQUILIBRIUM DYNAMICS }
\vspace*{0.035truein}
\centerline{\bf IN HOT QED PLASMAS}
\vspace*{0.37truein}
\centerline{\footnotesize SHANG-YUNG WANG}
\vspace*{0.015truein}
\centerline{\footnotesize\it Department of
Physics and Astronomy, University of Pittsburgh}
\baselineskip=10pt \centerline{\footnotesize\it
Pittsburgh, Pennsylvania 15260, U.S.A.}

%\vspace*{0.225truein}\publisher{(September 24, 2000)}{(October 16, 2000)}

\vspace*{0.21truein}
\abstracts{The quantum kinetics of photons in hot QED
plasmas is studied directly in real time
by implementing the dynamical renormalization group.
In contrast to conventional approach, the
dynamical renormalization group method consistently includes
off-shell (energy non-conserving) effects and accounts for
time-dependent collisional kernel. To lowest order we find that
in the relaxation time approximation the semihard photon distribution
function relaxes with a power law.}{}{}

\textlineskip          %) USE THIS MEASUREMENT WHEN THERE IS
\vspace*{12pt}         %) NO SECTION HEADING

The quantum kinetic description to study Abelian and non-Abelian
plasmas in extreme environments is of fundamental importance in
the understanding of the formation and evolution of a novel phase
of matter, the quark-gluon plasma (QGP), expected to be produced
in ultrarelativistic heavy ion experiments. The typical approach
to derive quantum kinetic equations begins by introducing a Wigner
transform of a nonequilibrium Green's function at two different
space-time points\cite{geiger} and often requires a quasiparticle
approximation.\cite{danielewicz} The rationale behind the Wigner
transform is the assumption of a wide separation between the
microscopic (fast) and the kinetic (slow) time scales, typically
justified in weakly coupled theories. Nevertheless, the
quasiparticle picture, closely related to the assumption of
completed collisions, is less warranted in ultrarelativistic heavy
ion collisions, as estimates based on energy deposited in the
central collision region at BNL RHIC energies $\sqrt{s}\sim
200\,A\;{\rm GeV}$ suggest that\cite{qgp} the lifetime of a
deconfined phase of quark-gluon plasma is of order $10-20\,{\rm fm}/c$
with an overall freeze-out time of order $100\,{\rm fm}/c$.

We consider a hot, neutral QED plasma which is prepared
at time $t=t_0$ with fermions in thermal equilibrium
at temperature $T$ but photons out of equilibrium.
The initial nonequilibrium photon distribution
functions are given by $n_{\bf k}(t_0)$.
The question that we want to ask is
{\em how does $n_{\bf k}(t)$ evolve in time?}
To make the calculation tractable
we will focus on {\em semihard} photons of momentum $eT\ll k\ll T$
($e\ll 1$ is the electromagnetic coupling constant), as the photon
self-energy for semihard photon is dominated by the hard thermal
loop\cite{htl} (HTL) contribution which is still perturbative.
%Furthermore, because of the abelian nature of electromagnetic interaction,
%we can work in a gauge invariant formulation in which physical
%observables are manifestly gauge invariant.\cite{boyanQED}

To find out the evolution of $n_{\bf k}(t)$ directly in
real time with initial condition $n_{\bf k}(t_0)$ specified,
we set up an {\em initial value problem} for $n_{\bf k}(t)$.
This can be achieved as follows.
First we construct a (Heisenberg) number operator $N_{\bf k}(t)$
that counts semihard photons of momentum ${\bf k}$
in terms of the (transverse) photon field operator and its conjugate
momentum, and find $dN_{\bf k}(t)/dt$ by using the Heisenberg equations of motion.
Then we express $dn_{\bf k}(t)/dt\equiv\langle dN_{\bf k}(t)/dt\rangle$
in terms of the nonequilibrium expectation value of the photon and fermion fields
in nonequilibrium field theory and compute perturbatively in powers
of $e$ using the nonequilibrium Feynman rules and real-time propagators.
To lowest order we find\cite{boyanQED}
\begin{eqnarray}
\frac{d}{dt}n_{\bf k}(t)&=&[1+n_{\bf k}(t_0)]
\,\Gamma^<_k(t)-n_{\bf k}(t_0)\,\Gamma^>_k(t),\label{photon:ndot2}\\
\Gamma^{\mbox{\scriptsize
\raisebox{1.8pt}{\raisebox{1.8pt}{$\scriptscriptstyle>$}
\raisebox{-1pt}{$\scriptscriptstyle\!\!\!\!\!\!<$}}}}_k(t)&=&
\int_{-\infty}^{+\infty} d\omega\,{\cal R}^{\mbox{\scriptsize
\raisebox{1.8pt}{\raisebox{1.8pt}{$\scriptscriptstyle>$}
\raisebox{-1pt}{$\scriptscriptstyle\!\!\!\!\!\!<$}}}}_k(\omega)\,
\frac{\sin[(\omega-k)(t-t_0)]}{\omega-k}.\label{Gamma}
\end{eqnarray}
Here $\Gamma^{\mbox{\scriptsize
\raisebox{1.8pt}{\raisebox{1.8pt}{$\scriptscriptstyle>$}
\raisebox{-1pt}{$\scriptscriptstyle\!\!\!\!\!\!<$}}}}_k(t)$ is
the {\em time-dependent} rates and
${\cal R}^{<(>)}_k(\omega)$ has a physical
interpretation in terms of the {\em off-shell}
(i.e., {\em energy non-conserving})
photon production (absorption) processes in the plasma:
bremsstrahlung $e^-(e^+)\leftrightarrow e^- (e^+)\gamma$ and
annihilation $e^- e^+ \leftrightarrow\gamma$.
Since the fermions are in thermal equilibrium,
the Kubo-Martin-Schwinger (KMS) condition
${\cal R}^>_k(\omega)=e^{\omega/T}\,
{\cal R}^<_k(\omega)$ holds.
In the HTL approximation a detailed analysis\cite{boyanQED} reveals that
${\cal R}^{\mbox{\scriptsize
\raisebox{1.8pt}{\raisebox{1.8pt}{$\scriptscriptstyle>$}
\raisebox{-1pt}{$\scriptscriptstyle\!\!\!\!\!\!<$}}}}_k(\omega)$
is completely determined by the off-shell (energy non-conserving)
processes $e^-(e^+)\leftrightarrow e^- (e^+)\gamma$.
%which is also responsible for Landau-damping.

It should be emphasized that in the {\em infinite time limit},
the resonance factor in Eq.~(\ref{Gamma}) can be approximated by
a delta function, {\it viz.},
$$
\frac{\sin[(\omega-k)(t-t_0)]}{\pi(\omega-k)}\buildrel{t\rightarrow
\infty}\over{\approx}\delta(\omega-k),
$$
which is exactly the assumption of completed collisions
invoked in time-dependent perturbation theory leading to
{\em energy conservation} and Fermi's golden rule
(i.e., a time-independent rate).
In this limit the above off-shell (energy non-conserving)
processes cannot take place due to kinematics,
and the kinetics of photons should arise from higher order contributions.

For any {\em finite} time, however,
Eq.~(\ref{photon:ndot2}) can be solved by
direct integration, we find
\begin{equation}
n_{\bf k}(t)=n_{\bf k}(t_0)+
\left[1+n_{\bf k}(t_0)\right]
\int^t_{t_0}dt'\,\Gamma^<_k(t')
-\,n_{\bf k}(t_0)\int^t_{t_0}dt'\,\Gamma^>_k(t').\label{ngamma2}
\end{equation}
The integrals that appear in the above expression,
$$
\int^t_{t_0}dt'\; \Gamma^{\mbox{\scriptsize
\raisebox{1.8pt}{\raisebox{1.8pt}{$\scriptscriptstyle>$}
\raisebox{-1pt}{$\scriptscriptstyle\!\!\!\!\!\!<$}}}}_k(t') =
\int_{-\infty}^{+\infty} d\omega\,{\cal R}^{\mbox{\scriptsize
\raisebox{1.8pt}{\raisebox{1.8pt}{$\scriptscriptstyle>$}
\raisebox{-1pt}{$\scriptscriptstyle\!\!\!\!\!\!<$}}}}_k(\omega)\,
\frac{1-\cos[(\omega-k)(t-t_0)]}{(\omega-k)^2},
$$
are dominated, in the long time limit, by the regions of
$\omega$ for which the denominator is resonant (i.e, $\omega\approx k$).
Using the HTL approximation for ${\cal R}^{\mbox{\scriptsize
\raisebox{1.8pt}{\raisebox{1.8pt}{$\scriptscriptstyle>$}
\raisebox{-1pt}{$\scriptscriptstyle\!\!\!\!\!\!<$}}}}_k(\omega)$,
we find that\cite{boyanQED}
\begin{equation}
\int^t_{t_0} dt'\, \Gamma^<_k(t') \buildrel{k(t-t_0) \gg 1}\over=
\frac{e^2 T^3}{6k^3}\left\{\ln\left[2k(t-t_0)\right]+\gamma-1\right\}+
{\cal O}[1/k(t-t_0)],
\label{detbalance}
\end{equation}
where $\gamma=0.577\ldots$ is the Euler's constant.

Do we find the time evolution of $n_{\bf k}(t)$? No, because
at large times the (logarithmic) {\em secular} term in Eq.~(\ref{detbalance})
that grows in time will invalidate the perturbative solution Eq.~(\ref{ngamma2}).
The next step of our kinetic approach is to resum these secular terms
by implementing the dynamical renormalization group,\cite{boyanQED,boyanrgk}
which introduces a renormalization of the distribution function via
$$
n_{\bf k}(t_0)={\cal Z}_k(t_0,\tau)n_{\bf k}(\tau),\quad
{\cal Z}_k(t_0,\tau)=1+e^2 z_k^{(1)}(t_0,\tau)+e^4
z_k^{(2)}(t_0,\tau)+\cdots.
$$
The renormalization coefficients $z^{(i)}_k(t_0,\tau)$ are
chosen to cancel the secular divergence order by order
at an {\em arbitrary} time $\tau$
(see Wang {\it et al.}\cite{boyanQED} for details).
The perturbative solution can be evolved to large times
provided that $\tau$ is chosen close to $t$.
A change in the time scale $\tau$ is compensated by a change
of the $n_{\bf k}(\tau)$ in such a manner that $n_{\bf k}(t)$
does not depend on the arbitrary scale $\tau$.
This independence leads to the
{\em dynamical renormalization group equation},
which consistently to order $e^2$ is given by\cite{boyanQED}
\begin{equation}
\frac{d}{d\tau}{n}_{\bf k}(\tau)=[1+n_{\bf k}(\tau)]
\,\Gamma^{<}_k(\tau)-
n_{\bf k}(\tau)\, \Gamma^{>}_k(\tau)+{\cal O}(e^4).\label{photon:drg}
\end{equation}
Choosing $\tau$ to coincide with $t$ in Eq.~(\ref{photon:drg}),
we obtain the {\em quantum kinetic equation}
\begin{equation}
\frac{d}{dt}{n}_{\bf k}(t)=[1+n_{\bf k}(t)]\;\Gamma^{<}_k(t)-
n_{\bf k}(t)\;\Gamma^{>}_k(t).\label{photon:keq}
\end{equation}

In the relaxation time approximation, which describes
the approach to equilibrium of a slightly off-equilibrium
mode with initial distribution $n_{\bf k}(t_0)=n_B(k)+\delta
n_{\bf k}(t_0)$ [$n_B(k)$ is the Bose-Einstein thermal distribution]
while all the other modes are in equilibrium,
solving Eq.~(\ref{photon:keq}) we find a power law relaxation
at large times to be given by\cite{boyanQED}
\begin{equation}
\delta n_{\bf k}(t)=\delta n_{\bf k}(t_0)\,[k(t-t_0)]^{-e^2 T^2/6k^2}
\quad\mbox{ for}\quad k(t-t_0)\gg 1.\label{rta}
\end{equation}

To conclude, our approach to derive kinetic equations in hot
plasmas is different from the one often used in the literature
which involves the Wigner transform and the assumption of completed
collisions. In particular, it reveals clearly the dynamics of off-shell
(energy non-conserving) effects arising from the finite system
lifetime. This aspect is of phenomenological importance in the
study of the {\em direct photon production} from a
quark-gluon plasma of finite lifetime.\cite{photon}

\vspace*{5pt}
\noindent {\em Acknowledgements}. The author would like to thank
D.\ Boyanovsky, H.\ J.\ de Vega, and D.-S.\ Lee for collaboration.
This work was supported in part by the US NSF through grants PHY-9605186
and PHY-9988720.

\end{document}